\documentclass[superscriptaddress,reprint,amsmath,amssymb,aps,floatfix,longbibliography]{revtex4-1}

\usepackage{amsmath,gensymb,textcomp,bm,dcolumn,eurosym,array,tabu,multirow,nicefrac,color,subfigure,graphicx,upgreek}
\usepackage[colorlinks, linkcolor=blue, citecolor=blue, urlcolor=blue, breaklinks=true]{hyperref}

\usepackage{mathpazo,times} 

\newcommand{\ket}[1]{\lvert #1 \rangle}
\newcommand{\braket}[2]{\langle #1 \vert #2 \rangle}

\usepackage[subsectionbib]{bibunits}
\usepackage{graphicx}
\usepackage{dcolumn}
\usepackage{bm,MnSymbol}
\usepackage{units}
\usepackage{psfrag}
\usepackage{float}
\usepackage{color}

\begin{document}

\title{Hong-Ou-Mandel interferometry on a biphoton beat note}

\author{Yuanyuan Chen}
\email{chenyy@smail.nju.edu.cn}
\affiliation{
Institute for Quantum Optics and Quantum Information - Vienna (IQOQI), Austrian Academy of Sciences, Boltzmanngasse 3, 1090 Vienna, Austria.}
\affiliation{Vienna Center for Quantum Science \& Technology (VCQ), Faculty of Physics, University of Vienna, Boltzmanngasse 5, 1090 Vienna, Austria}
\affiliation{State Key Laboratory for Novel Software Technology, Nanjing University, Xianlin Avenue 163, Nanjing 210046, China.}

\author{Matthias Fink}
\affiliation{
Institute for Quantum Optics and Quantum Information - Vienna (IQOQI), Austrian Academy of Sciences, Boltzmanngasse 3, 1090 Vienna, Austria.}
\affiliation{Vienna Center for Quantum Science \& Technology (VCQ), Faculty of Physics, University of Vienna, Boltzmanngasse 5, 1090 Vienna, Austria}

\author{Fabian Steinlechner}
\email{Fabian.Steinlechner@iof.fraunhofer.de}
\affiliation{Fraunhofer Institute for Applied Optics and Precision Engineering IOF, Albert-Einstein-Strasse 7, 07745 Jena, Germany.}
\affiliation{Friedrich Schiller University Jena, Abbe Center of Photonics, Albert-Einstein-Str. 6, 07745 Jena, Germany.}

\author{Juan P. Torres}
\affiliation{
ICFO-Institut de Ciencies Fotoniques, The Barcelona Institute of Science and Technology, 08860 Castelldefels (Barcelona), Spain}
\affiliation{Department of Signal Theory and Communications, Universitat Politecnica de Catalunya, 08034 Barcelona, Spain}

\author{Rupert Ursin}
\email{Rupert.Ursin@oeaw.ac.at}
\affiliation{
Institute for Quantum Optics and Quantum Information - Vienna (IQOQI), Austrian Academy of Sciences, Boltzmanngasse 3, 1090 Vienna, Austria.}
\affiliation{Vienna Center for Quantum Science \& Technology (VCQ), Faculty of Physics, University of Vienna, Boltzmanngasse 5, 1090 Vienna, Austria}


\begin{abstract}
Hong-Ou-Mandel interference, the fact that identical photons that arrive simultaneously on different input ports of a beam splitter bunch into a common output port, can be used to measure optical delays between different paths. It is generally assumed that great precision in the measurement requires that photons contain many frequencies, i.e., a large bandwidth. Here we challenge this ``well-known'' assumption and show that the use of two well-separated frequencies embedded in a quantum entangled state (discrete color entanglement) suffices to achieve great precision. We determine optimum working points using a Fisher Information analysis and demonstrate the experimental feasibility of this approach by detecting thermally-induced delays in an optical fiber. These results may significantly facilitate the use of quantum interference for quantum sensing, by avoiding some stringent conditions such as the requirement for large bandwidth signals.
\end{abstract}


\maketitle

\section{Introduction}
The exploitation of quantum interference promises to enhance sensing technologies beyond the possibilities of classical physics. Hong-Ou-Mandel (HOM) interference is a prototypical example of such a quantum phenomenon, that lacks any counterpart in classical optics. When two identical photons in a global pure state impinge on a beam splitter from separate input modes, they both leave the beam splitter through the same output port, as a consequence of their bosonic nature \cite{hong1987measurement}. On the other hand, if the input photons are not identical, or they are independent but not in a pure state, the ``bunching'' probability is directly related to the photons' level of indistinguishability, or its degree of purity \cite{mosley2008heralded}. This effect enables a wide range of quantum information processing tasks, ranging from the characterization of ideally identical single-photon emitters \cite{aharonovich2016solid}, the implementation of photonic Bell state measurements for entanglement swapping and quantum teleportation \cite{pan2012multiphoton}, or in tailoring high-dimensional entangled states of light \cite{zhang2016engineering,ndagano2019entanglement}.

HOM interferometry also holds great promise for sensing schemes that require precise knowledge of optical delays. When the relative arrival time of two photons is varied, the coincidence rate exhibits a characteristic dip with a width that is related to the photons' coherence time. Notably, and unlike other interferometric approaches based on first-order interference, HOM interference is not affected by variations in the optical phase. As a consequence, a HOM interferometer maintains its ability to measure time delays, even when fluctuations of path length difference are on the order of the wavelength. This feature has resulted in proposals for HOM-based time delay sensors with an ultrahigh timing resolution \cite{lyons2018attosecond} and novel protocols such as Quantum Optical Coherence Tomography (QOCT) that benefit from other quantum features, such as the cancellation of some deleterious dispersion effects \cite{nasr2003demonstration}.

In the context of such applications, the broad consensus has been that the width of the dip, i.e. the coherence time, imposes the ultimate limit on the precision. As a consequence, ultra-broad-band photon sources have long been hailed as a vital prerequisite for ultra-precise HOM interferometry.

Here we embark on an alternative route towards ultra-precise HOM interferometry using superpositions of two well-separated and entangled discrete frequency modes and coincidence detection on the bi-photon beat note. The manifestation of this fourth-order spatial beating effect is an oscillation within a typically Gaussian envelope that is determined by the coherence time of the two photons, as a direct result of relative phase shift between distinct colors \cite{ou1988observation,rarity1990two}. We explore the sensitivity limits as a function of the difference frequency of color-entangled states, as imposed by the Quantum Cram$\acute{e}$r-Rao (QCR) bound, and find that the precision with which the delays can be measured is mainly determined not by the coherence time of photons, but by the separation of the center frequencies of the state. Aside from promising improved precision, the approach allows to increase the dynamic range of a HOM-based sensor, provided the required frequency non-degenerate states can be generated in a tunable manner.

We show how suitable frequency entangled states are readily obtained with comparatively little technological effort by employing a variation of the source scheme recently developed in Ref. \cite{chen2018polarization}. Building on the measurement and estimation strategy by analyzing the Fisher information to determine the optimum working points for frequency-degenerate HOM interference, recently proposed in Ref. \cite{lyons2018attosecond}, we experimentally demonstrate an optimized HOM sensor that we use to detect delays introduced by temperature drifts in an optical fiber.

The results obtained in this proof of concept experiment show that quantum interference of unconventional frequency states on a beam splitter provides a simple way of enhancing the timing resolution in HOM-based sensors and may also indicate a new direction towards fully harnessing HOM interference in quantum sensing and quantum information processing.

\section{Results}
\subsection{HOM interfereometry with frequency entangled states}
Let us search for the ultimate limits to the precision of a HOM-based sensor. We consider the generic task of estimating an unknown parameter $\tau$ of a physical system. We prepare a probe state $\ket{\Psi_0}$ that is transformed as $\ket{\Psi_0}\rightarrow\ket{\Psi(\tau)}$ upon interaction with the physical system. The transformed state is then subjected to a particular measurement strategy to obtain an estimator of $\tau$. Irrespective of the specifics of the final measurement step, we may already state a fundamental limit for the precision of estimation $\delta \tau$ \cite{helstrom1969quantum,fujiwara1995quantum}:
\begin{equation}
\begin{split}
\delta\tau\geq \frac{1}{2\sqrt{NQ}}=\delta\tau_{QCR},
\end{split}
\end{equation}
where
\begin{equation} \label{eq:q}
\begin{split}
Q=\langle\frac{\partial\Psi(\tau)}{\partial\tau}|\frac{\partial\Psi(\tau)}{\partial\tau}\rangle-|\langle\Psi(\tau)|\frac{\partial\Psi(\tau)}{\partial\tau}\rangle|^2
\end{split}
\end{equation}
and $N$ is the number of independent trials of the experiment. The generality of this statement, known as the Quantum Cram$\acute{e}$r-Rao bound, is remarkable: no matter what ingenious measurement procedure the experimenter may contrive, she will never achieve a precision better than $\delta \tau_{QCR}$. Since the QCR bound is attached to a particular quantum state, it is clear that the appropriate choice of the probe state is of the utmost importance \cite{pirandola2018advances}.

Let us now consider an experimental configuration where paired photons (signal and idler), with central frequencies $\omega_1^0$ and $\omega_2^0$, originate from a parametric down-conversion process (SPDC) pumped by a CW pump with frequency $\omega_p=\omega_1^0+\omega_2^0$. Each photon of the pair is injected into one of the two arms of a HOM interferometer. The time delay of interest is one that may occur due to an imbalance between the two arms of the interferometer. Even though the common case in HOM interferometry is to consider signal and idler photons with the same central frequency, in the following we allow for a more general configuration where the state of interest is a discrete or continuous frequency entangled state \cite{ramelow2009discrete}:
\begin{equation}
\begin{split}
\ket{\Psi(\tau)}=&\frac{1}{\sqrt{2}}\int d\Omega f(\Omega) \times \\
&[e^{i(\Delta+2\Omega) \tau}a_1^\dag(\omega_1^0+\Omega)a_2^\dag(\omega_2^0-\Omega) - \\
&a_1^\dag(\omega_2^0+\Omega)a_2^\dag(\omega_1^0-\Omega)]\ket{vac},
\end{split}
\end{equation}
where $\Delta=\omega_1^0-\omega_2^0$ is the difference frequency of two well-separated center frequency bins, $\ket{vac}$ is the vacuum state, and $f(\Omega)$ is an the Gaussian spectral amplitude function with $\int d\Omega|f(\Omega)|^2=1$. For this state, the Quantum Cram$\acute{e}$r-Rao limit on the estimation of time delays writes
\begin{equation}\label{eq:quantum cramer rao bound}
\begin{split}
\delta\tau_{QCR}= \frac{1}{N^{1/2}}\frac{1}{(\Delta^2+4\sigma^2)^{1/2}},
\end{split}
\end{equation}
where $\sigma=\sqrt{\langle\Omega^2\rangle-\langle\Omega\rangle^2}$ is the RMS (root mean square) bandwidth of SPDC photons. The dependence of the QCR bound on frequency detuning $\Delta$ gives us a first indication to the potential use of non-degenerate frequency entanglement as an alternative to large bandwidth for enhanced resolution HOM interferometry.

Up until now, we have only considered limitations that are inherent to the particular choice of the quantum state. We must confirm that we can experimentally realize this potential benefit using an appropriate measurement strategy, i.e. one that allows us to saturate equation \eqref{eq:quantum cramer rao bound}. As we shall see in the following, this can be accomplished via coincidence detection in the output ports of a balanced beam splitter. The beam splitter transforms the bi-photon state (see Methods for details) to
\begin{equation}
\ket{\Psi(\tau)}\rightarrow\ket{\Psi_A(\tau)}+\ket{\Phi(\tau)},
\end{equation}
where $\ket{\Psi_A(\tau)}$ and $\ket{\Phi(\tau)}$ correspond to the events that two photons emerge in opposite and identical outports, respectively. The normalized coincidence detection probability $P_c(\tau)=|\braket{\Psi(\tau)}{\Psi_A(\tau)}|^2$ reads
\begin{equation}\label{eq:interference probability}
\begin{split}
P_c(\tau)=\frac{1}{2}[1+ cos(\Delta\tau+\phi)exp(-2\sigma^2\tau^2)],
\end{split}
\end{equation}
where $\phi$ is a relative phase factor.

In the case of a real HOM interferometer, that is subject to photon loss $\gamma$ and imperfect experimental visibility $\alpha$, there are three possible measurement outcomes; either both photons are detected, one photon is detected, or no photon detected. The corresponding probability distributions read
\begin{equation}\label{eq:loss probability}
\begin{split}
P_2(\tau)&=\frac{1}{2}(1-\gamma)^2[1+\alpha cos(\Delta\tau)e^{-2\sigma^2\tau^2}]\\
P_1(\tau)&=\frac{1}{2}(1-\gamma)^2[\frac{1+3\gamma}{1-\gamma}-\alpha cos(\Delta\tau)e^{-2\sigma^2\tau^2}]\\
P_0(\tau)&=\gamma^2,
\end{split}
\end{equation}
where subscripts 0, 1 and 2 denote the number of detectors that click, corresponding to total loss, bunching and coincidence, respectively. For a more detailed discussion refer to \cite{lyons2018attosecond}. The outcome probabilities in this measurement can now be used to construct an \emph{estimator} for the value of $\tau$.

An estimator $\tilde{\tau}$ is a function of the experimental data that allows us to infer the value of the unknown time delay using a particular statistical model for the probability distribution of measurement outcomes. It is thus itself a random variable, that can be constructed from the probability distributions $P_i(\tau)$ as a function of time delay. The average of an \emph{unbiased estimator} corresponds to the real time delay. For any such estimator, classical estimation theory states standard deviation is lower bounded by
\begin{equation}\label{eq:cramer rao bound}
\begin{split}
\delta\tau_{CR}= \frac{1}{(N F_\tau)^{1/2}} \geq \delta\tau_{QCR},
\end{split}
\end{equation}
where the Fisher information $F_\tau$ reads
\begin{equation}
\label{eq: Fisher information}
F_\tau=\frac{(\partial_\tau P_2(\tau))^2}{P_2(\tau)}+\frac{(\partial_\tau P_1(\tau))^2}{P_1(\tau)}+\frac{(\partial_\tau P_0(\tau))^2}{P_0(\tau)}.
\end{equation}
This limit is known as the Cram$\acute{e}$r Rao bound. It is tied to a particular quantum state and a specific measurement strategy. Evaluating the Fisher information for this set of probabilities, we find that its upper bound is achieved in ideal case ($\gamma=0$, $\alpha=1$) at position of $\tau\rightarrow0$ as
\begin{equation}
\label{eq: maximal Fisher information}
\begin{split}
\lim_{\tau\rightarrow0}F_\tau=\Delta^2+4\sigma^2.
\end{split}
\end{equation}
In the case of zero loss and perfect visibility we recover the Quantum Cram$\acute{e}$r Rao Bound, thus confirming that the measurement strategy is indeed optimal.

While equation \eqref{eq:cramer rao bound} provides an ultimate bound on the achievable precision of estimation that can be achieved, the approach does not yet tell us how to construct a suitable estimator for $\tau$. To this end a widely used analytical technique is maximum likelihood estimation (MLE). The likelihood function $\mathcal{L}(\tau)$ is defined from measurement outcomes, whose logarithm can be maximized by using optimization algorithm such as Gradient Descent to predict the parameter $\tau$ that we want to infer. In our framework, the likelihood function is a multinomial distribution as $\mathcal{L}(N_0,N_1,N_2|\tau)\propto P_0(\tau)^{N_0}P_1(\tau)^{N_1}P_2(\tau)^{N_2}$, where $N_0$, $N_1$ and $N_2$ denote the numbers of events that no, only one and two detector(s) click(s), respectively. Note that $P_0(\tau)$, being independent of $\tau$, results in a constant scale factor that is of no relevance to the final calculation of the Fisher information and parameter estimation. The likelihood is extremized as \cite{lyons2018attosecond}:
\begin{equation}
\begin{split}
0&=:(\partial_\tau log\mathcal{L})_{\tilde{\tau}_{MLE}}\\
&=\frac{N_0P_0(\tau)^\prime}{P_0(\tau)}+\frac{N_1P_1(\tau)^\prime}{P_1(\tau)}+\frac{N_2P_2(\tau)^\prime}{P_2(\tau)},
\end{split}
\end{equation}
and solving this equation enables us to predict an optimal estimator as $\tilde{\tau}_{MLE}$.

\subsection{Experiment}
\textbf{Experimental realization of bi-photon beat note.}
\begin{figure}[!t]
\centering
\includegraphics[width=\linewidth]{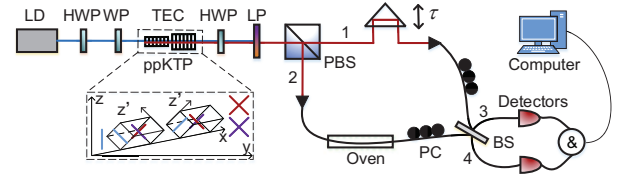}
\caption{Experimental setup for temperature sensor through beating of frequency entanglement. LD: laser diode; HWP: half wave plate; WP: wave plate; ppKTP: type-II periodically poled potassium titanyl phosphate crystal; TEC: temperature controller; LP: long pass filter; PBS: polarizing beam splitter; Oven: computer-controlled heating device; PC: polarization controller; BS: beam splitter; Detectors: single photon detectors.}
\label{figure_1}
\end{figure}
We generate photon pairs via spontaneous parametric down-conversion pumped with a continuous-wave pump laser. The experimental configuration implemented to generate the desired frequency entangled state of distant frequency modes (i.e. signal and idler frequencies that are separated by more than the spectral bandwidth $\omega_{s}-\omega_{i}\gg \Delta \omega$) is a modified crossed-crystal configuration \cite{kwiat1999ultrabright,steinlechner2012high} shown in the inset of Fig.\ \ref{figure_1}. In this configuration, two nonlinear crystals for type-II SPDC are placed in sequence, whereby the optical axis of the second crystal is rotated by 90$^\circ$ with respect to the first. Balanced pumping of the two crystals ensures equal probability amplitudes for SPDC emission $\ket{V,\omega_p}\rightarrow \ket{V,\omega_s}\ket{H,\omega_i}$ in the first-, or $\ket{H,\omega_p}\rightarrow \ket{H,\omega_s}\ket{V,\omega_i}$ in the second crystal, where $H/V$ denote horizontal and vertical polarizations. The photons are guided to a PBS, which maps the orthogonally polarized photon pairs into two distinct spatial modes ($1,2$) in the desired frequency entangled state
\begin{equation}\label{eq:frequency state}
\begin{split}
\ket{\psi}_\omega^- \rightarrow 
(\frac{\ket{\omega_s}_{1}\ket{\omega_i}_{2} - \ket{\omega_i}_{1}\ket{\omega_s}_{2}}{\sqrt{2}})\otimes\ket{H}_1\ket{V}_2.
\end{split}
\end{equation}

The frequency entangled photons are routed to the input ports of a beam splitter. After operation of HOM interference, we only focus on the situation that two detectors indiscriminately register coincidence, i.e., exiting via different ports, as a direct consequence of anti-bunching effect of photons entangled in the form of anti-symmetric state.

\begin{figure}[!t]
\centering
\includegraphics[width=\linewidth]{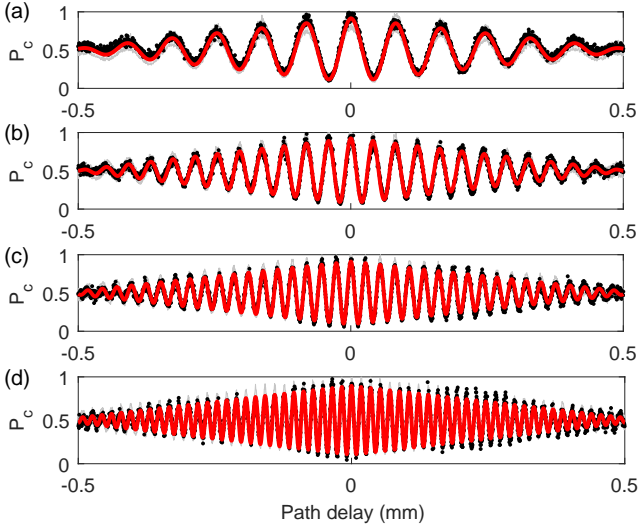}
\caption{ Two photon HOM interference of frequency entanglement with different frequency detunings of (a) $\unit[3.65]{THz}$ ($\unit[7.98]{nm}$) at temperature of $\unit[30]{^\circ C}$, (b) $\unit[7.35]{THz}$ ($\unit[16.08]{nm}$) at temperature of $\unit[50]{^\circ C}$, (c) $\unit[11.18]{THz}$ ($\unit[24.45]{nm}$) at temperature of $\unit[70]{^\circ C}$ and (d) $\unit[17.08]{THz}$ ($\unit[37.35]{nm}$) at temperature of $\unit[100]{^\circ C}$.}
\label{figure_2}
\end{figure}
As the central wavelengths of down-converted photons are related to the phase-matching temperature of nonlinear crystals, our source has the ability to produce color tunable frequency entangled photon pairs. We analyze the HOM signal for various frequency detunings to demonstrate this flexibility (see Fig.\ \ref{figure_2}). By fitting these interference fringes to normalized coincidence probability as equation \eqref{eq:interference probability}, we are able to estimate single photon frequency bandwidth to be $\unit[0.253]{THz}$, which corresponds to a bandwidth in wavelength of $\unit[0.55]{nm}$ and a coherence time of $\unit[3.5]{ps}$. Frequency detunings are much larger than single photon bandwidth such that two frequency bins could be separated completely. The visibilities of these experimentally measured frequency entangled photon pairs can reach $0.85\pm0.05$. The maximal frequency detuning we have measured is $\unit[17.08]{THz}$ at temperature of $\unit[100]{^\circ C}$, which is about 68 times the single photon frequency bandwidth.

\textbf{Fisher information analysis.}
\begin{figure*}[!t]
\centering
\includegraphics[width=\linewidth]{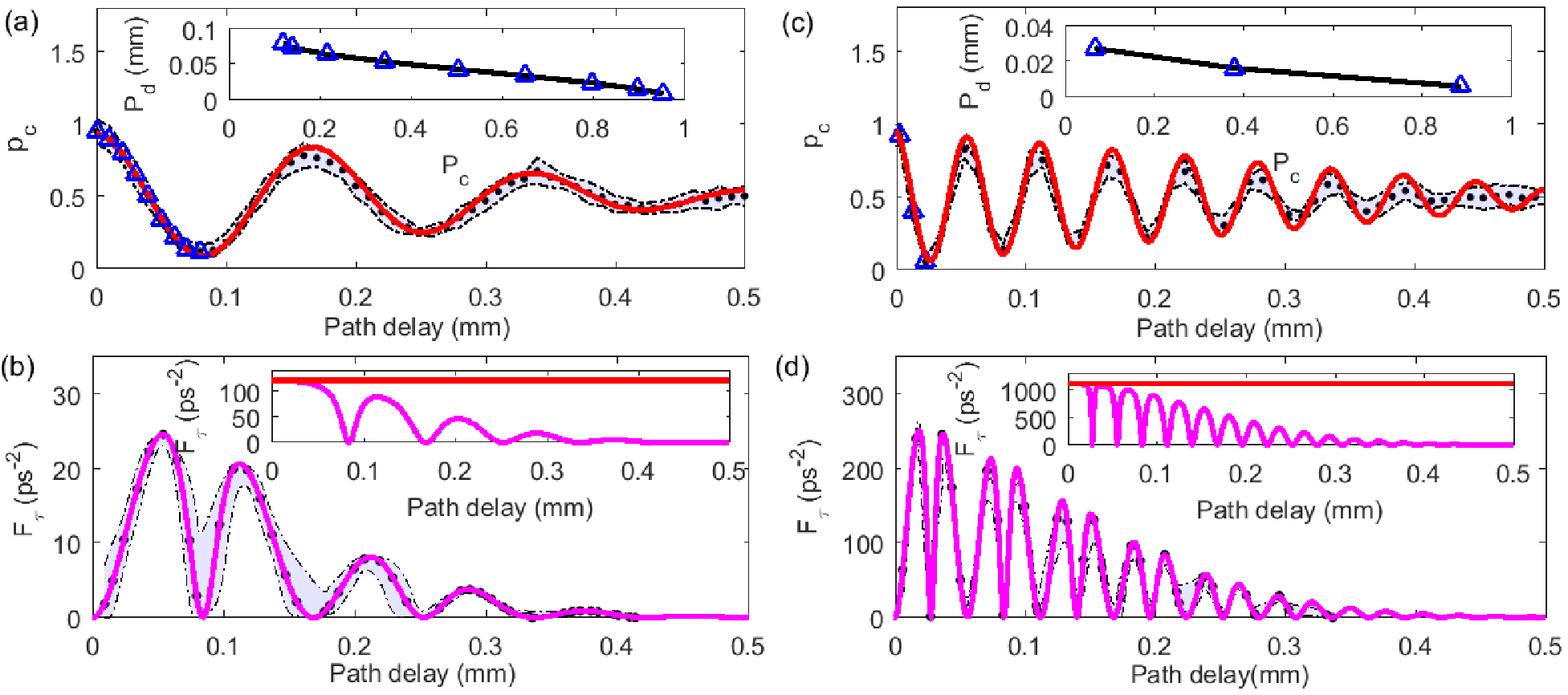}
\caption{Experimental description of Fisher information. Hong-Ou-Mandel interference of frequency entanglement with detunings of (a) \unit[1.75]{THz} and (c) \unit[5.34]{THz}, where the insets show the relative path delay predicted from normalized coincidence probability (for simplicity, only estimation results of blue data points are shown). These parameters enable us to calculate Fisher information in experiments for frequency entanglement with detunings of (b) \unit[1.75]{THz} and (d) \unit[5.34]{THz} ($\gamma = 0.4$ and $\alpha = 0.9$), where the insets show their theoretical simulation in ideal case ($\gamma = 0$ and $\alpha = 1$), and the red lines represent the ultimate limits of achievable Fisher information calculated from Cram$\acute{e}$r-Rao bound. The shaded regions bounded by two smoothed curves represent the standard deviation of experimental results estimated by statistical methods assuming a
Poisson distribution.}
\label{figure_3}
\end{figure*}
Figure \ref{figure_3} demonstrates the explicit procedure of parameter estimation and their corresponding Fisher information in experiment, from which we see that frequency detuning can facilitate the achievement of higher resolution and precision. The oscillation of Fisher information within two-photon coherence time is a key signature of discrete frequency entanglement \cite{ramelow2009discrete}. Here the maximal Fisher information we have obtained is $\unit[245]{ps^{-2}}$ for frequency detuning of $\unit[5.34]{THz}$, which inversely reveals the highest precision of $\unit[639]{as}$, i.e., relative path delay of $\unit[192]{nm}$, for experimental trials of $O(10^4)$. It is noticed that the quadratic dependence of Fisher information as a function of frequency detuning could be used to further enhance the Fisher information with respect to the frequency degenerate case, where values of $\sim \unit[8]{ps^{-2}}$ have already been reported \cite{lyons2018attosecond}.

\textbf{Experimental application as a temperature sensor.}
\begin{figure*}[!t]
\centering
\subfigure[]{
\label{Fig4.sub.1}
\includegraphics[width=0.49\linewidth]{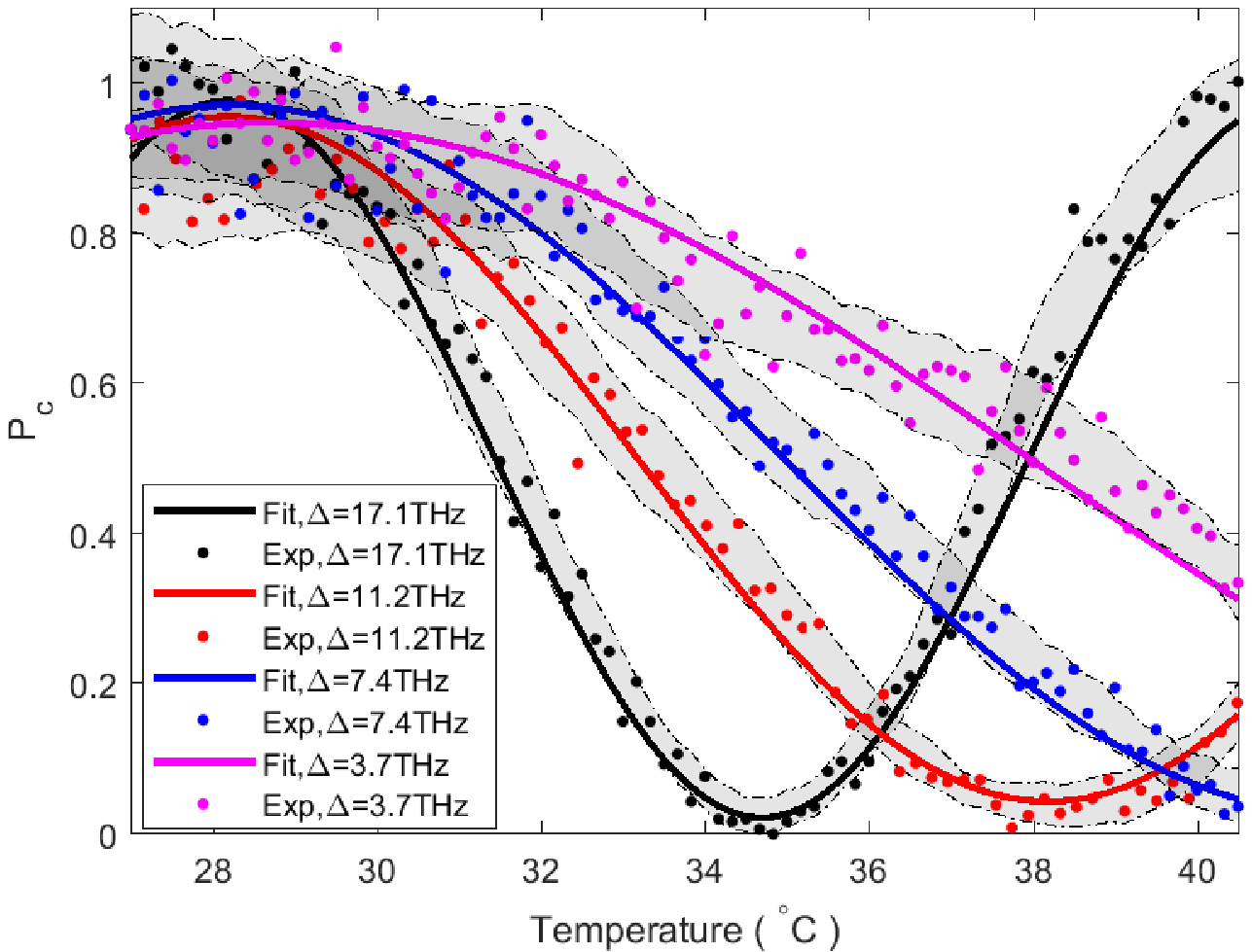}}
\subfigure[]{
\label{Fig4.sub.2}
\includegraphics[width=0.49\linewidth]{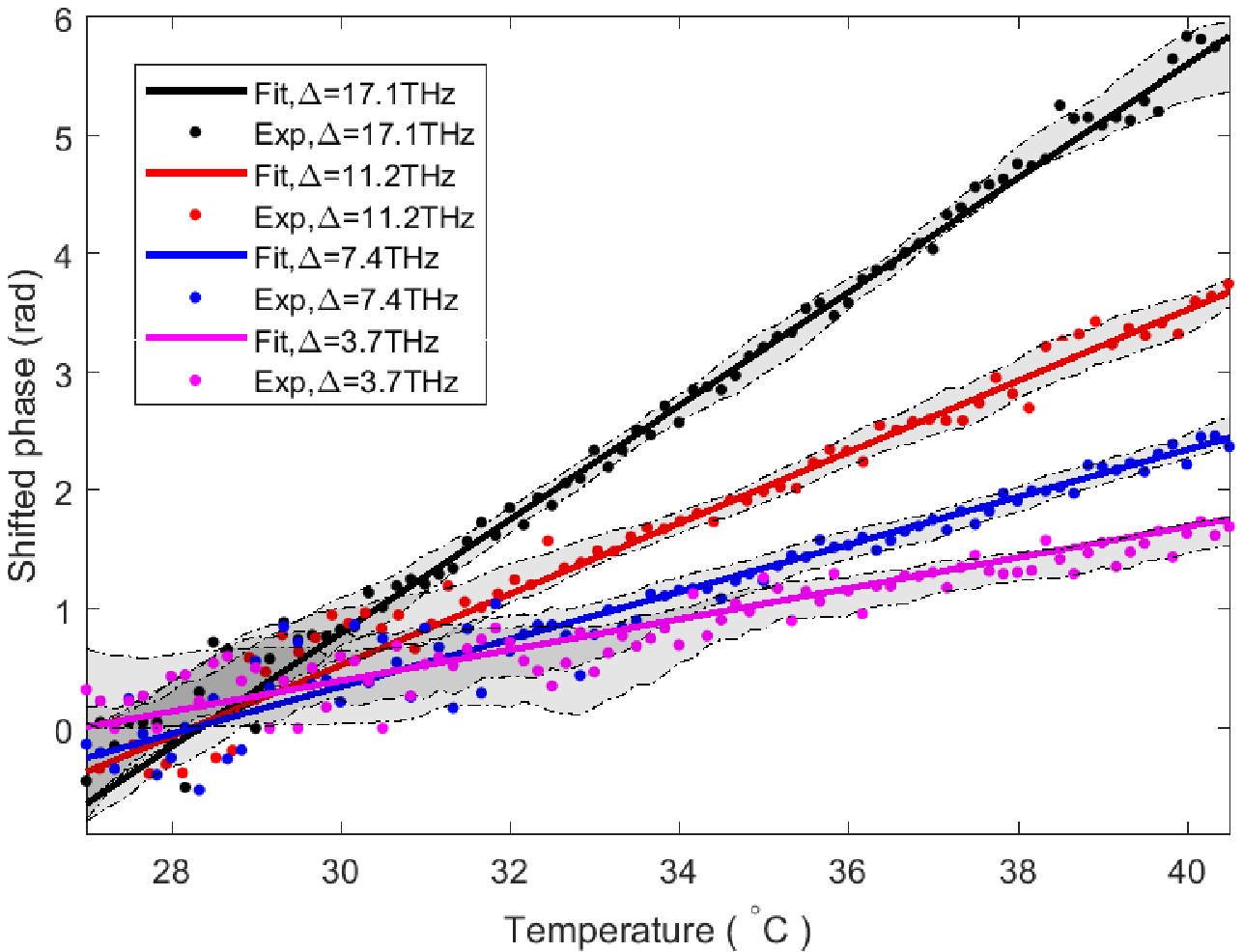}}
\caption{Experimental demonstration of thermal characteristics of jacket optical fiber. (a) Two-fold coincidence probability and (b) corresponding shifted phase as a function of heating temperature of sensing fiber versus different frequency detunings. The shaded regions bounded by two smoothed curves represent the standard deviation of experimental results estimated by statistical methods assuming a Poisson distribution.}
\label{figure_4}
\end{figure*}
In order to demonstrate the viability principle of employing our HOM sensor, we performed a proof of concept experiment in which we estimate the time delay due to linear expansion of a jacket optical fiber.
In order to verify the conclusion that quantum metrology based on frequency entanglement with larger frequency detuning has higher precision, we experimentally measure two-fold coincidence probabilities and predict the thermal coefficient by heating the sensing fiber to vary relative phase shifts (see Fig.\ \ref{figure_4}).

In principle, the relative phase shift varies almost linearly with fiber length and is described as $\beta=N_gkL$, where $L$ is sensing fiber length, $N_g$ is the material group index and $k$ is the light wave number \cite{lagakos1981temperature}. Since the input frequency entangled state of HOM sensor is highly sensitive to transmission time, the relative phase shift, introduced by the length extension of fiber, can be expressed as a function of heating temperature, and resulting in the thermal coefficient as
\begin{equation}\label{eq:coefficient}
\frac{d\beta}{dT}\approx(\frac{2\pi}{\lambda_s}-\frac{2\pi}{\lambda_i})\frac{dN}{dT}L_o+(\frac{2\pi}{\lambda_s}N_o^{\lambda_s}-\frac{2\pi}{\lambda_i}N_o^{\lambda_i})\frac{dL}{dT},
\end{equation}
where $\lambda_{s/i}$ is center wavelength of signal or idler photons, $N_o^{\lambda_{s/i}}$ and $L_o$ are the corresponding parameters at room temperature, $T$ is heated temperature, $\frac{dN}{dT}$ and $\frac{dL}{dT}$ are thermal coefficients of material group index and fiber length, respectively.

We notice that the thermal coefficient of shifted phase is related to frequency detuning, which agrees well with the experimental measurement results (see Fig.\ \ref{Fig4.sub.2}), and results in the coincidence probability varies as cosine function (see Fig.\ \ref{Fig4.sub.1}). The measured thermal coefficients is $\unit[0.13]{rad/deg}$, $\unit[0.2]{rad/deg}$, $\unit[0.3]{rad/deg}$ and $\unit[0.48]{rad/deg}$ for frequency detunings of $\unit[3.7]{THz}$, $\unit[7.4]{THz}$, $\unit[11.2]{THz}$ and $\unit[17.1]{THz}$, respectively. The refractive index of pure silica is wavelength dependent, and its first derivative with respect to temperature is about $\unit[1\times10^{-5}]{/deg}$ \cite{bruckner1970properties}. Then we are able to estimate the thermal coefficient of linear expansion of jacket optical fiber to be $\frac{dL}{dT} \sim \unit[4.8\times10^{-7}]{m/deg}$, which agrees well with the results reported in Ref. \cite{tateda1980thermal,priest1997thermal}. Accordingly the maximal frequency detuning that we observed in this proof-of-principle experiment enables us to achieve temperature resolution of $\unit[0.12]{deg}$.

\section{Discussion}
We have demonstrated a new approach to HOM interferometry based on discrete frequency entanglement of well separated frequency modes and detection of a beat note coincidence signal.

Previous HOM-interferometric sensing schemes required perfect frequency degenerate and ultra-broad-band SPDC emission. Any wavelength distinguishability decreases visibility of the HOM dip and correspondingly diminishes the resolution. Providing suitable quantum sources for this case is a significant challenge, as it either requires the engineering of aperiodic poling structures or the use of very short nonlinear crystals, at the cost of efficiency. In contrast, the approach outlined here requires only a sufficiently large non-degeneracy, whereby the spectral bandwidth can be small. We have experimentally demonstrated how to generate suitable discrete frequency-entangled states, in a manner that can be readily extended to larger wavelength separations. For example, $\lambda_s$= $\unit[1,500]{nm}$ and $\lambda_2$ = $\unit[800]{nm}$ ($\unit[1,000]{THz}$ angular difference frequency) a timing sensitivity of 9 as could already be obtained for only $N = 10^4$ detection events. Backed by the results of our proof-of-concept experiment, this shows that the approach can provide higher resolution and highly sensitive measurement, and makes it an ideal candidate for more quantum enhanced metrology applications.

Although this work only reports the advantages of our approach in estimating delays, similar great enhancement can also be achieved for a variety of applications like state discrimination or hypothesis testing.

In conclusion, we believe that fully harnessing HOM interference and frequency entanglement will provide  additional tools, e.g. for frequency shaping of photons and interference phenomena in general, ultimately broadening the path towards practical quantum applications.
\\

\newpage
\section{Methods}
\textbf{Entangled Photon source.}
In our experimental realization of flexible frequency entanglement source \cite{chen2018polarization}, two mutually orthogonally oriented 10-mm-long ppKTP crystals are manufactured to provide collinear phase matching with pump (p), signal (s) and idler (i) photons at center wavelengths of $\lambda_p \approx \unit[405]{nm}$ and $\lambda_{s,i} \approx \unit[810]{nm}$. They are pumped with a $\unit[405]{nm}$ continuous wave grating-stabilized laser diode. To achieve the desired diagonally and anti-diagonally polarization states for simplifying alignment, we design an oven with V-groove such that two crossed crystals are oriented at $\unit[45]{^\circ}$. Since the pump beam is horizontally polarized, it is equally likely to generate a photon pair in the first or second crystal, resulting in a state of equation \eqref{eq:frequency state}. The relative phase factor is compensated by tilting a half wave plate. Long pass filter is set to block pump beam. Then PBS routes a pair of photons into two distinct spatial modes according to orthogonal polarizations.

\textbf{HOM interferometer.}
In spatial mode 1, a translation stage introduces a relative path delay to accomplish the task of scanning HOM interference fringes. Polarization controllers are required to compensate polarization difference of biphotons such that only frequency entanglement can make contributions to the interference effect. Finally the anti-bunched photons are detected by silicon avalanche photon diodes, and two-fold events are identified using a fast electronic AND gate when two photons arrive at the detectors within a coincidence window of $\sim \unit[3]{ns}$.

\textbf{Coincidence signal with frequency-entangled states.}
An optimal measurement procedure may allow us to saturate the limit set by equation \eqref{eq:cramer rao bound}. In the ideal, lossless with perfect visibility, case, such a measurement can be accomplished by interference on a balanced beam splitter. The beam splitter transformation on the input modes can be expressed by
\begin{equation}
\begin{split}
\hat{a}_1^\dag(\omega_1)=\frac{1}{\sqrt{2}}[\hat{a}_3^\dag(\omega_1)+\hat{a}_4^\dag(\omega_1)]\\
\hat{a}_2^\dag(\omega_2)=\frac{1}{\sqrt{2}}[\hat{a}_3^\dag(\omega_2)-\hat{a}_4^\dag(\omega_2)],
\end{split}
\end{equation}
where $\omega_1$ and $\omega_2$ denote the signal or idler frequency mode that are incident from opposite ports, and subscripts 1/2 (3/4) represent two input (output) ports of that beam splitter. Accordingly the state is transformed as
\begin{equation}
\ket{\Psi(\tau)}\rightarrow\ket{\Psi_A(\tau)}+\ket{\Phi(\tau)},
\end{equation}
where these state contributions can be expressed as
\begin{equation}
\begin{split}
\ket{\Psi_A(\tau)}=&\frac{1}{2}\int d\Omega f(\Omega)(1+e^{-i\Delta\tau})\\
&[\hat{a}_3^\dag(\omega_1^0+\Omega)\hat{a}_4^\dag(\omega_2^0-\Omega)\\
&-\hat{a}_3^\dag(\omega_2^0-\Omega)\hat{a}_4^\dag(\omega_1^0+\Omega)]\ket{vac}\\
\ket{\Phi(\tau)}=&\frac{1}{2}\int d\Omega f(\Omega)(1-e^{-i\Delta\tau})\\
&[\hat{a}_3^\dag(\omega_1^0+\Omega)\hat{a}_3^\dag(\omega_2^0-\Omega)\\
&-\hat{a}_4^\dag(\omega_1^0+\Omega)\hat{a}_4^\dag(\omega_2^0-\Omega)]\ket{vac}.
\end{split}
\end{equation}
Due to HOM interference on the beam splitter coincidence detection in distinct spatial modes projects onto the state component $\ket{\Psi_A(\tau)}$.

\textbf{Fisher information.}
In a specific experiment (measurement strategy), with outcomes $x_i$, and corresponding probability distributions $P_i(\tau )$, any unbiased estimator will fulfill equation \eqref{eq:cramer rao bound}, where the Fisher information $F_\tau$ quantifies the information that a particular measurement can reveal about the unknown parameter of interest. Note that optimizing over all probability distributions results we recover the QCR bound. The outcomes of this measurement are sufficient to obtain an estimator for the value of $\tau$.

By substituting equation \eqref{eq: Fisher information} with the corresponding probabilities from equation \eqref{eq:loss probability}, we could calculate the Fisher information as
\begin{equation}
\begin{split}
F_\tau=\frac{(1-\gamma^2)[\alpha\Delta sin(\Delta\tau)+4\alpha\sigma^2\tau cos(\Delta\tau)]^2e^{-4\sigma^2\tau^2}}{4P_1(\tau)P_2(\tau)/(1-\gamma)^4}.
\end{split}
\end{equation}
We note that the Fisher information is undefined at position of $\tau = 0$ in ideal case since the denominator will be zero.

\textbf{Maximum-likelihood estimator.} Since no prior knowledge is provided, we can apply maximum likelihood estimation approach to predict the target parameter. We extremized the likelihood function as
\begin{equation}
\begin{split}
0&=:(\partial_\tau log\mathcal{L})_{\tilde{\tau}_{MLE}}=\frac{N_1P_1(\tau)^\prime}{P_1(\tau)}+\frac{N_2P_2(\tau)^\prime}{P_2(\tau)}.
\end{split}
\end{equation}
Based on the calculation in equation \eqref{eq:loss probability}, we know $P_1(\tau)^\prime=-P_2(\tau)^\prime$ such that
\begin{equation}
N_1P_2(\tau)|_{\tilde{\tau}_{MLE}}=N_2P_1(\tau)|_{\tilde{\tau}_{MLE}}.
\end{equation}
For the sake of simplicity, $\tilde{\tau}$ in term of $e^{-2\sigma^2\tau^2}$ can be considered as a constant value, i.e., coarse sensing position $\tau_s$ where Fisher information is highest. Thus we get an optimal estimator to variable relative time delay as
\begin{equation}
\begin{split}
\tilde{\tau}_{MLE}=\arccos(\frac{\frac{1+3\gamma}{1-\gamma}N_2-N_1}{\alpha(N_1+N_2)e^{-2\sigma^2\tau_s^2}})/\Delta,
\end{split}
\end{equation}
and the values of parameters $\tau_s$, $\gamma$, $\alpha$, $\sigma$ and $\Delta$ need to be separately estimated before the measurements begin.

\textbf{Shifted phase of temperature sensor.}
The introduced phase shift in frequency entangled state can be expressed as a function of heating temperature as
\begin{equation}\label{eq:linear phase}
\begin{split}
\beta=&\frac{2\pi}{\lambda_s}(\frac{dN}{dT}L_o+\frac{dL}{dT}N_o^{\lambda_s}+\frac{dN}{dT}\frac{dL}{dT}T)T\\
&-\frac{2\pi}{\lambda_i}(\frac{dN}{dT}L_o+\frac{dL}{dT}N_o^{\lambda_i}+\frac{dN}{dT}\frac{dL}{dT} T)T.
\end{split}
\end{equation}
Since $\frac{dN}{dT}\frac{dL}{dT}$ is in the order of much smaller magnitude, we ignore the term of $\frac{dN}{dT}\frac{dL}{dT}T$ in equation \eqref{eq:linear phase}.

\textbf{See supplementary materials for more information.}
\section*{Data availability}
Data available on request from the authors.

\section*{Acknowledgements}
We thank Thomas Scheidl, Sebastian Ecker, Soeren Wengerowsky, Johannes Handsteiner, Siddarth Joshi, and Lukas Bulla for experimental support and helpful conversations. YC thanks Lijun Chen for support. The research leading to these results has received funding from the H2020 European Programme under Grant Agreement 801060 Q-MIC, the Austrian Research Promotion Agency (FFG) Projects - Agentur f\"{u}r Luft- und Raumfahrt (FFG-ALR contract 6238191 and 866025), the European Space Agency (ESA contract 4000112591/14/NL/US) as well as the Austrian Academy of Sciences. YC acknowledges personal funding from Major Program of National Natural Science Foundation of China (No. 11690030, 11690032), National Key Research and Development Program of China (2017YFA0303700); the National Natural Science Foundation of China (No.61771236), and from a Scholarship from the China Scholarship Council (CSC) and the program B for Outstanding PhD candidate of Nanjing University. This work was supported by the Fraunhofer Internal Programs under Grant No. Attract 066-604178. JPT acknowledges financial support from Fundacio Cellex, from the Government of Spain through the Severo Ochoa Programme for Centres of Excellence in R\&D (SEV-2015-0522), from Generalitat de Catalunya under the programs ICREA Academia and CERCA, and from the project 17FUN01 BeCOMe within the Programme EMPIR, and initiative co-founded by the European Union and the EMPIR Participating Countries.

\section*{Contributions}

F.S. developed the initial idea for this work. Y.C. conducted the experiment under supervision from F.S. and R.U. Theoretical analysis was carried out by J.T. and Y.C.. Y.C. and F.S. wrote the first draft and all authors contributed to the final version of the manuscript.

\section*{Competing interests}

The authors declare that there are no competing interests.

\bibliography{apssamp}
\bibliographystyle{naturemag}
\end{document}